# Optical Dipole Trapping beyond Rotating Wave Approximation: The case of Large Detuning

**Vassilios E. Lembessis[1] and Demosthenes Ellinas[2]**

[1] *New York College, 38 Amalias Str., GR 105 58, Athens, Greece*

(vlembessis@yahoo.com)

[2] *Department of Sciences, Division of Mathematics, Technical University of Crete,*

*GR 731 00, Chania, Crete, Greece*  (ellinas@science.tuc.gr)

**Abstract**

We show that the inclusion of counter-rotating terms, usually dropped in evaluations of interaction of an electric dipole of a two level atom with the electromagnetic field, leads to significant modifications of trapping potential in the case of large detuning. The results are shown to be in excellent numerical agreement with recent experimental findings, for the case of modes of Laguerre-Gauss spatial profile.

## I. INTRODUCTION

It is well known that the interaction of laser light with atoms leads to cooling, heating and trapping of atoms [1, 2, 3]. The characteristics of the trapping potential depend on a number of parameters, most notably the detuning ($\Delta = \omega_0 - \omega_L$) where $\omega_0$ is the atomic transition frequency in a two-level atomic model and $\omega_L$ is the laser frequency. The potential associated with the dipole force, in the case of small detuning ($\Delta \ll \omega_0$) is given by,

$$U = U(\mathbf{R}) = \frac{\hbar\Delta}{2}\ln\left(1 + \frac{2\Omega^2(\mathbf{R})}{\Delta^2 + \Gamma^2}\right), \qquad (1)$$

where Γ is the line width of the atomic excited state and Ω(**R**) the interaction Rabi frequency.

The above result has emerged from a theory in which the counter-rotating terms have been dropped. In this work we focus our attention on the case of very large detuning. In theoretical works carried out in the field of laser cooling and trapping little attention has been given to those counter-rotating terms, which usually are omitted from the interaction Hamiltonian [4], but in cases of very large detuning these terms become important so our work is devoted to a theoretical estimation of their magnitudes. Based on their relative magnitude these terms are found to be a negligible correction to experiment such that the one reported by Chu et. al. in [5], but they give a significant contribution and in fact agree with numerical estimations done for an experiment such that the one done by Stamper-Kurn et. al. and reported in [6].

## II. THE DIPOLE FORCE REVISED

We wish to examine the average dipole force acting on a two-level atom or ion in the case of very large detuning where the counter-rotating terms in the interaction Hamiltonian become very important and thus cannot be neglected. In this case we have the so called Far Off-Resonance Traps (FORT) which are currently under investigation [7]. An advantage of the optical dipole force is that confines atoms in all hyperfine levels so the two-level assumption is still valid for large detuning.

We follow a method based on Heisenberg's operators perturbation techniques which has been successfully used elsewhere [8, 9]. The light is in the form of a coherent beam with a complex amplitude $a$ and with a generic spatial distribution [8]. We consider $a$, $a^+$ as the creation and annihilation operators which satisfy the boson commutation property, $[a, a^+] = 1$. The beam is considered as linearly polarised. The choice of such a polarization has been considered as the right one because the magnetic sub-levels $m_F$ of a certain hyperfine ground state $F$ are shifted by the same amounts. This only requires a detuning large compared to the excited-state hyperfine splitting which is very well fulfilled in dipole trapping experiments. If the detuning also exceeds the ground state hyperfine splitting then all sub states of the electronic ground state are equally shifted, and the dipole potential becomes independent of $m_F$ and $F$ [10].

The quantum mechanical description of the two-level atom is given by kets spanning a state-space of two dimensions. Thus we consider the ground state $|1\rangle$ and the excited state $|2\rangle$ separated by an amount of energy $\hbar\omega_0$. The operators that excite and de-excite the two-level atom can be written as $\pi^+ = |2\rangle\langle 1|$ and $\pi = |1\rangle\langle 2|$ and they satisfy the anti-commutation relation, $\pi\pi^+ + \pi^+\pi = 1$, for all common times. The appropriate Hamiltonian is given by

$$H = H_A + H_F + H_{INT} \tag{2}$$

where $H_A$ and $H_F$ represent the Hamiltonian for the atom and field, given by $H_A = \mathbf{P}^2/2M + \hbar\omega_0 \pi^+\pi$ and $H_F = \hbar\omega_L a^+ a$ respectively. With $\mathbf{P}$ we denote the centre-of-mass momentum operator. The interaction Hamiltonian $H_{INT}$ describes the coupling of the atom with the electromagnetic field assuming electric dipole approximation and is given by

$$H_{INT} = -\mathbf{d} \cdot \mathbf{E}(\mathbf{R}) \tag{3}$$

where **d** is the atomic dipole moment operator and $\mathbf{E}(\mathbf{R})$ is the electric field evaluated at the position **R** of the atom. We can use the excitation and de-excitation operators in order to obtain the quantum mechanical expression for the atomic dipole moment which corresponds to the transition between the ground and the excited state. Indeed the dipole moment operator is given by

$$\mathbf{d} = \mathbf{D}_{12}(\pi^+ + \pi) = \mathbf{D}_{12}(|2\rangle\langle 1| + |1\rangle\langle 2|) \tag{4}$$

where $\mathbf{D}_{12} = \langle 1|\mathbf{d}|2\rangle = \langle 2|\mathbf{d}|1\rangle$ and we assume that the wave-functions corresponding to the atomic ground and excited states are real [11]. The electric field vector associated with a mode propagating along the z-axis is given by

$$\mathbf{E}(\mathbf{R}) = i\{\vec{\varepsilon}\, a\, H(\mathbf{R})\exp(i\Theta(\mathbf{R})) - h.c\} \tag{5}$$

where $\vec{\varepsilon}$ is the polarization vector in the $x-y$ plane, H(**R**) is the amplitude of the field which in general depends on the position as well as the phase Θ(**R**) of the field. The term $h.c$ stands as usual for the hermitian conjugate quantity. The coupling between the atom and field is given by

$$H_{int} = -\mathbf{d}.\mathbf{E}(\mathbf{R}) = -ih[\pi^+ a G(\mathbf{R})e^{i\Theta(\mathbf{R})} - h.c] - ih[\pi a G(\mathbf{R})e^{i\Theta(\mathbf{R})} - h.c] \tag{5}$$

with $G(\mathbf{R}) = (\mathbf{D}_{12} \cdot \vec{\varepsilon})H(\mathbf{R})/h$. The terms in the second bracket represent the non energy-conserving terms, this means that we do not proceed in the rotating wave approximation.

The time evolution of the system is derivable from the Heisenberg equation of motion. For an operator *O* this is

$$d O/ dt = \frac{1}{ih}[O, H] \tag{6}$$

which can be formally integrated to give

$$O(t) = O(0) + \frac{1}{ih}\int_0^t [O(t'), H(t')]dt' \tag{7}$$

where $O(0)$ denotes the initial time value of $O$. We now obtain the time evolution for the atomic and field annihilation operators given by the following respective expressions,

$$\pi(t') = \exp(-i\omega_0 t')\pi(0) + G(\mathbf{R})\int_0^{t'} dt''[2\pi^+(t'')\pi(t'')-1][a(t'')\exp(i\Theta(\mathbf{R})-i\omega_0(t'-t''))] -$$

$$G(\mathbf{R})\int_0^{t'} dt''[2\pi^+(t'')\pi(t'')-1]a^+(t'')\exp(-i\Theta(\mathbf{R})-i\omega_0(t'-t''))], \tag{8}$$

$$a(t') = \exp(-i\omega_L t')a(0) + G(R)\int_0^{t'} dt'' \exp(-i\Theta(R)-i\omega_L(t'-t''))\pi(t'') +$$

$$G(R)\int_0^{t'} dt'' \exp(-i\Theta(R)-i\omega_L(t'-t''))\pi^+(t''). \tag{9}$$

We may use the notation

$$\langle \mathbf{P}(t) \rangle = \langle \Psi | \mathbf{P}(t) | \Psi \rangle \tag{10}$$

to denote the expectation value of the momentum operator in a well defined state $|\Psi\rangle$ which is only characterised by the initial average photon number $n_\mathbf{k}$ and occupation numbers $n_g$, $n_e$ of the ground and excited state of the two-level atom respectively then we set

$$n_\mathbf{k} = \langle \Psi | a^+(0)a(0) | \Psi \rangle, n_g = \langle \Psi | \pi(0)\pi^+(0) | \Psi \rangle, n_e = \langle \Psi | \pi^+(0)\pi(0) | \Psi \rangle. \tag{11}$$

The occupation numbers for the ground and excited state, respectively, satisfy the equation $n_e + n_g = 1$, which expresses the fact that atomic transitions occur strictly between the excited and the ground state. We may assume $n_e = 0$, as well as $\langle \Psi | \pi(0) | \Psi \rangle = \langle \Psi | \pi^+(0) | \Psi \rangle = 0$, initially. We assume also that $a(0)|\psi\rangle = |a|e^{i\phi}|\psi\rangle$ thus $\langle \psi | a^2(0) | \psi \rangle = |a|^2 e^{i2\phi} = n_k e^{i2\phi}$. We must now make some important remarks. At $t = 0$ we consider that the atom is at $\mathbf{R}$ and we also consider the detuning $\Delta = \omega_0 - \omega_L$ and the Rabi frequency given by $\Omega(\mathbf{R}) = \sqrt{n_k} G(\mathbf{R})$. A new parameter that will play important role in our results is the sum of the two frequencies involved defined by

$$Z = \omega_0 + \omega_L. \tag{12}$$

The operator for the atomic centre-of-mass momentum is $\mathbf{P} = -i\hbar \vec{\nabla}_{\mathbf{R}}$ where $\mathbf{R}$ is taken to be the atomic centre-of-mass position. The time dependent average value of the momentum operator is given by

$$\langle \mathbf{P}(t') \rangle = \mathbf{P}(0) + \left\langle \psi \left| (i/\hbar) \int_0^t dt' [H(t'), \mathbf{P}(t')] \right| \psi \right\rangle. \tag{13}$$

The evaluation of this quantity is a bit laborious and, throughout the calculation we retain terms only up to second order in the coupling parameter, that is second order in $G$. We present here the final result omitting phase gradients terms proportional to $\vec{\nabla}_{\mathbf{R}} \Theta(\mathbf{R})$ which are attributed to scattering force and become important near resonance. Then we obtain

$$\langle \mathbf{P}(t) \rangle = -\hbar \Omega(\mathbf{R}) \vec{\nabla}_{\mathbf{R}} (\Omega(\mathbf{R})) \left\{ \left[ -\frac{2 \sin \Delta t}{\Delta^2} + \frac{2t}{\Delta} + \frac{2 \sin Z t}{Z^2} - \frac{2t}{Z} \right] \right.$$
$$\left. + 4 \operatorname{Re} \left[ e^{i2\phi} e^{i2\Theta(\mathbf{R}(0))} \left[ i \frac{\cos(\Delta - Z)t}{Z(\Delta - Z)} + \frac{\sin(\Delta - Z)t}{Z(\Delta - Z)} + \frac{t}{Z} \right] \right] \right\}. \tag{14}$$

The average force on the atom $\langle \mathbf{F} \rangle$ associated with the atomic momentum $\langle \mathbf{P} \rangle$ is formally obtainable by direct use of the Heisenberg equation

$$\mathbf{F}(t) = \frac{d\mathbf{P}}{dt} = \frac{1}{i\hbar} [\mathbf{P}, H]. \tag{15}$$

This agrees with the classical assignment

$$\langle \mathbf{F}(t) \rangle = \frac{d}{dt} \langle \mathbf{P}(t) \rangle. \tag{16}$$

The expectation value is on both the internal degrees of freedom and on the center-of-mass state of the atom. We must point out that the temporal scale of the internal dynamics is determined by the spontaneous emission rate as well as the Rabi frequency, while the dynamics of center-of-mass has temporal scale determined by $h/E_R$ where $E_R$ is the recoil energy. In many cases these two frequency scales can be extremely different. If that is the case, the internal state of atoms can be assumed to be in a quasi-steady state relative to that of the center-of-mass, and the internal and external contributions to the force of Eq. (15) may be factorized. It is important to keep in mind that this factorisation scenario, which assumes in particular the absence of any quantum entanglement between the internal and center-of-mass motion of the atom is well justified in general. It becomes questionable for very strong Rabi frequencies. The gradient of the Rabi frequency (for example in Eq. (17)) may be a very serious problem if we have, for instance, ultra cold atoms. For well localised particles, however, one can roughly approximate this wave function by a $\delta$-function located at some position $\mathbf{R}(t=0)$.

From Eqs. (15), (16) we get the following expression for the dipole force

$$\langle \mathbf{F}(t) \rangle = -h\Omega(\mathbf{R})\vec{\nabla}_R(\Omega(\mathbf{R}))\left\{ \frac{4}{\Delta}\sin^2\left(\frac{\Delta t}{2}\right) - \frac{4}{Z}\sin^2\left(\frac{Zt}{2}\right) \right.$$
$$\left. + 4\,\mathrm{Re}\left[ e^{i2\varphi} e^{i2\Theta(\mathbf{R}(0))}\left[ -i\frac{\sin(\Delta-Z)t}{Z} + \frac{\cos(\Delta-Z)t}{Z} + \frac{1}{Z} \right] \right] \right\}. \tag{17}$$

The formalism so far has dealt with the interaction of the atom with only one radiation mode and account needs to be taken for the effects of vacuum modes, i.e. for the spontaneous emission from the excited state to the ground state. **A full treatment of such a case can be done only through the Optical Bloch Equations [11]. However when we use Heisenberg's operators perturbation techniques, as in our case, we may introduce the effects of spontaneous emission in a** phenomenological way as follows [8, 9]. Denoting the decay rate of the excited state by $\Gamma$, we must point out that for a relatively intense beam with a Rabi frequency larger than $\Gamma$ the decay rate of the excited state is modified to $\Gamma' = 2\left(\frac{\Gamma^2}{4} + \frac{\Omega^2(\mathbf{R})}{2}\right)^{1/2}$ [11]. If we define the availability of the excited state as the survival rate $dp/dt$ at time $t$ by $dp/dt = \Gamma' e^{-\Gamma' t}$, then the force is due to transition from the excited state the time averaged force is $\langle \overline{\mathbf{F}} \rangle = \int \langle \mathbf{F}(t) \rangle \Gamma' e^{-\Gamma' t} dt$, and by virtue of Eq. (17) is obtained to be

$$\langle \overline{\mathbf{F}} \rangle = -2\hbar \Omega(\mathbf{R}) \vec{\nabla}_R (\Omega(\mathbf{R})) \left\{ \frac{\Delta}{\Delta^2 + \Gamma^2 + 2\Omega^2(R)} - \frac{Z}{Z^2 + \Gamma^2 + 2\Omega^2(R)} \right.$$
$$\left. + 2\operatorname{Re}\left[ e^{i2\varphi} e^{i2\Theta(\mathbf{R}(0))} \left[ -i\frac{\Gamma(\Delta-Z)}{Z[\Gamma^2 + 2\Omega^2(R) + (\Delta-Z)^2]} + \frac{\Gamma^2 + 2\Omega^2(R)}{Z[\Gamma^2 + 2\Omega^2(R) + (\Delta-Z)^2]} + \frac{1}{Z} \right] \right] \right\}.$$

(18)

This force is of conservative character and it should be emphasized that its second and third terms are new, and that the third term is obviously far smaller than the first and second ones and then can be omitted. In such a case the resulting force is derived from the potential

$$U = \frac{\hbar\Delta}{2} \ln\left(1 + \frac{2\Omega^2(\mathbf{R})}{\Delta^2 + \Gamma^2}\right) - \frac{\hbar Z}{2} \ln\left(1 + \frac{2\Omega^2(\mathbf{R})}{Z^2 + \Gamma^2}\right)$$
$$- 2\hbar \operatorname{Re}\left[ e^{i2\varphi} e^{i2\Theta(\mathbf{R}(0))} \left[ \ln\left(1 + \frac{2\Omega^2(\mathbf{R})}{(\Delta-Z)^2 + \Gamma^2}\right) \left( i\frac{\Gamma(\Delta-Z)}{2Z} + \frac{(\Delta-Z)^2}{2Z} \right) - \frac{2\Omega^2(\mathbf{R})}{Z} \right] \right].$$

(19)

The right hand side of this equation contains in addition to the first term of the potential two more terms modifying the effective potential function and these are due to the non rotating wave approximation adopted in our considerations. In the case of very large detuning all these terms in the potential above are of comparable size. As a matter of fact we can demonstrate the relative size of each term by making an estimation based on parameters used in recent experiments in light induced atom trapping. Specifically for the parameters that were used in the experimental work reported by Chu et al [5], the first term in the potential is $10^4$ times larger than the second one, and $10^7$ times larger than the third one . So there is no need to take into account the counter-rotating terms in the Hamiltonian.

In the work reported by Stamper-Kurn et al [6], it is clearly stated that the 25% of the trapping is due to such terms so they must be taken into account. More explicitly in that experimental work the values of the parameters used were, $w_0 = 6 \mu m$, $\Gamma = 10 MHz$, $\Omega = 3,67 \times 10^2 \Gamma$, $\omega_0 = 3.2 \times 10^{15} r/s$, $\omega_L = 1.9 \times 10^{15} r/s$, $\Delta = 1.3 \times 10^{15} r/s$, $Z = 5 \times 10^{15} r/s$. **If we employ these values in our case and further nullify the values of phase $e^{i2\varphi}e^{i2\Theta(\mathbf{R}(0))}$, appearing in last equation, we deduce from the last equation that all new terms give important contributions. The act of nullifying the phase is generally valid, since the dependence of the binding potential on this absolute phase can always be removed, by using a rotating frame in the space of atomic state vectors (i. e performing an appropriate unitary transformation), which will shift the origin of the phase angles.**

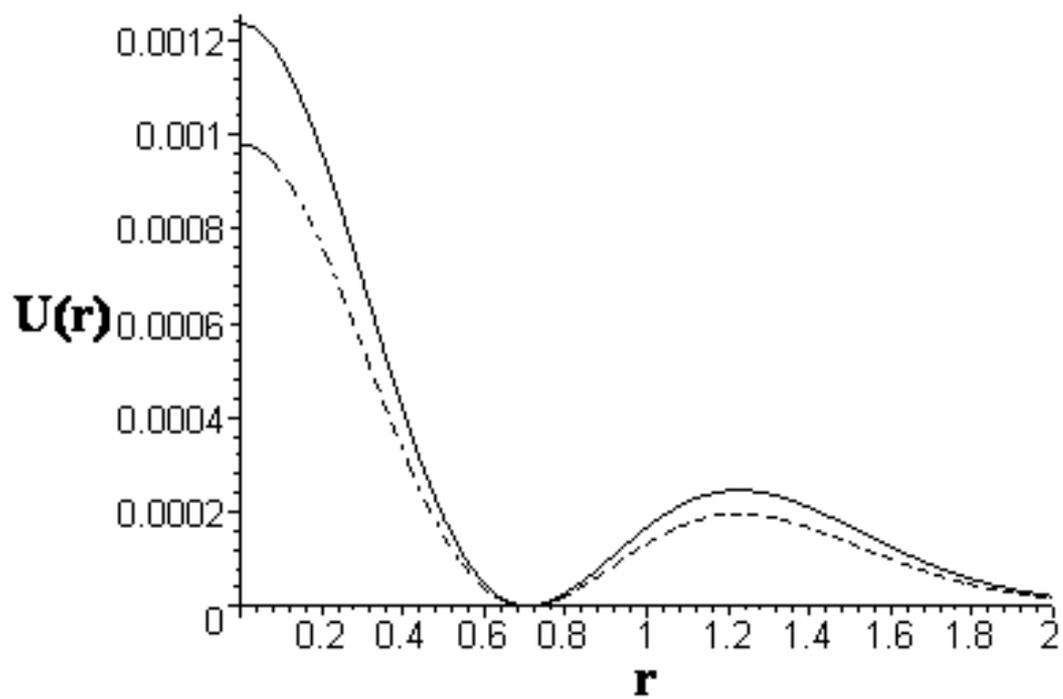

Figure 1a.

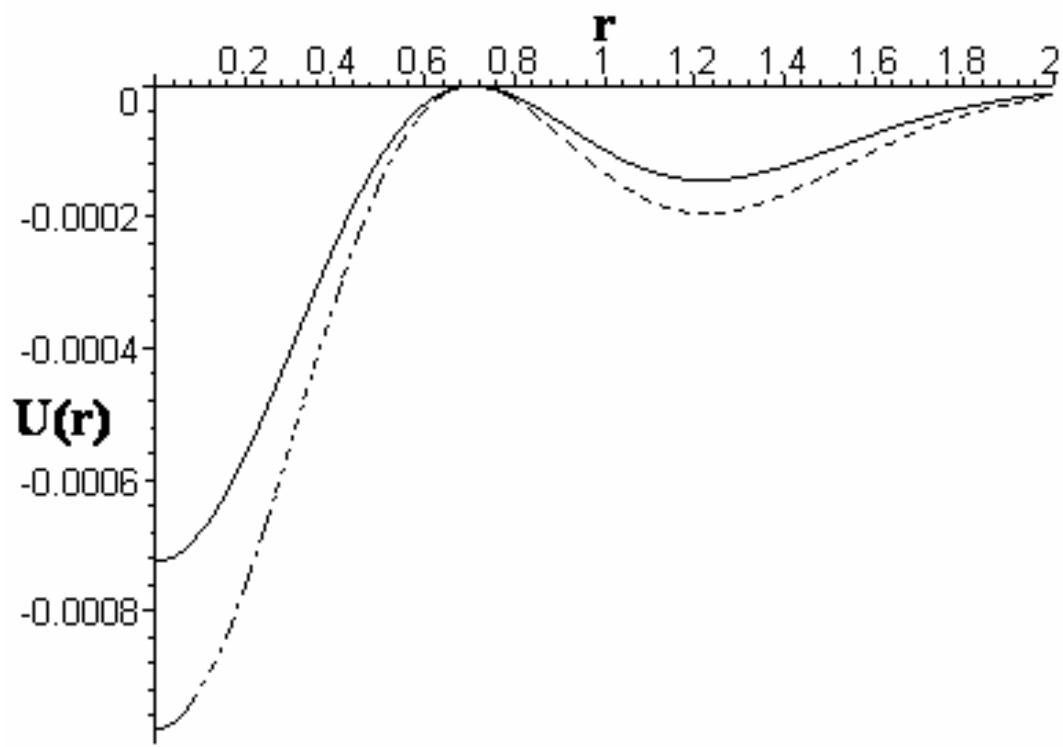

Figure 1b.

*Optical dipole potential when $\phi = 0$ for the case of positive (1a) and negative (1b) detuning. Our mode has the spatial profile of a Laguerre-Gaussian mode with l=0, p=2. In figure 1a the new terms lead to a higher potential (solid line) for positive detuning while in figure 1b (negative detuning) they lead to a shallower potential (solid line). In both figures the "RWA" potential is represented by the dashed line. The potentials are expressed in $h\Gamma$ units while the distance r is in $w_0$ units.*

Next in figures 1a and 1b, we present the trapping potentials for the cases of $\phi = 0$, in the non RWA (full line) and RWA (broken line) cases, respectively, for the Laguerre-Gaussian laser mode with *l=0, p=2* [12, 13].

We see that this choice of parameters give for positive detuning $\Delta > 0$, an effective trapping potential *higher* in the non RWA case that in RWA case ($U_{nonRWA} > U$), and reversibly for negative detuning $\Delta < 0$, a trapping potential for the non RWA case *more shallow* than in the RWA case ($U_{nonRWA} < U$, absolutely).

The difference in the potentials among the respective non RWA and RWA cases, is seen, either by direct numerical evaluation or by inspection of the values in the figures, to be about ¼. This is in excellent agreement with the estimation of the trapping potential reported in [6].

**VI. DISCUSSION**

By an analytic calculation we have shown first that in optical dipole traps the far off resonance condition causes a breaking of the rotating wave approximation which makes the contribution of counter rotating terms important in the potential which correspond to the optical dipole force by a factor of the order of 25%, for ranges of parameters used in current experiments. There is a very important point when we introduce the non-RWA terms in our calculations. Although the detuning term $\Delta$ may be either positive or negative, the sum of the frequencies given term Z is always positive. The presence of the later term may be either a contribution towards better or worse optical dipole trapping.

It is worthwhile to observe that in the case of positive detuning (e.g in Fig. 1a) where the trapping of atomic particles takes place in an area of high intensity which implies that the rate of spontaneous emission is high and therefore by recoil effects the decaying atoms may escape from the trap, the deepening of the trap in the non-RWA regime comes to a partial counterbalancing of this effect. In short, we have an improvement of trapping due to the breaking of RWA. On the contrary, the fact that in the case of negative detuning (e.g in Fig. 1b) where the trapping takes place in a zero intensity "dark" region where the spontaneous emission rate is negligible, the fact that the non-RWA terms give rise to a shallowing of the potential depth will not eventually affect the trapping effectiveness of our trapping scheme.

## Acknowledgments
The authors are grateful to Prof. M. Babiker for his interest and critical comments on the manuscript. We also thank Prof. W. Ketterle for providing us with experimental parameters. The two anonymous referees are thanked for their critical remarks.
This work is supported by "Pythagoras II" of EPEAEK research programm.